\documentclass[english,aps,prd,nofootinbib,showkeys,preprintnumbers,floatfix]{revtex4}
\usepackage[T1]{fontenc}
\usepackage[latin9]{inputenc}
\usepackage{graphicx}

\providecommand{\tabularnewline}{\\}

\usepackage{babel}

\begin{document}

\title{Searching Unparticle Signatures Through Tau Pair Production}

\author{O. Çak\i{}r}

\email{ocakir@science.ankara.edu.tr}

\affiliation{Department of Physics, Ankara University, 06100 Tandogan, Ankara,
Turkey}

\author{K. O. Ozansoy}

\email{oozansoy@science.ankara.edu.tr}

\affiliation{Department of Physics Engineering, Ankara University, 06100 Tandogan,
Ankara, Turkey}
\begin{abstract}
In this work, we study the possible unparticle signatures through
tau pair production $e^{+}e^{-}\to\tau^{+}\tau^{-}$ process at the
low energy particle factories. We take into account lepton flavor
conserving (s-channel) and violating (t-channel) couplings for the
unparticle leptonic interactions. For different values of the scaling
parameter $d$ we extract upper limits on the scalar unparticle couplings
at the integrated luminosities of 100 fb$^{-1}$/yr and 1 ab$^{-1}$/yr. 
\end{abstract}

\pacs{12.60.-i, 11.25.Hf, 14.80.-j, 13.15.+g}

\keywords{Unparticle sector, particle annihilation, new physics limits}

\maketitle

\section{Introduction}

High energy particle colliders that especially ones with multi-TeV
center of mass energies have indisputable importance to seek for and
to discover unknown physics effects. In many respects, the Large Hadron
Collider (LHC), the strongest particle collider ever built, is expected
to be launched in a year. There are several expectations to find some
outcomes for better understanding the physics of the electroweak scale
of the Standard Model (SM), and beyond it. To discover the Higgs particle,
if any, or to distinguish the effects due to the possible new physics
effects beyond the SM, such as supersymmetric models, or extra dimensions,
are main goals of the prospected gigantic collider.

Beside those high energy colliders, the rather low energy colliders
have been running for precise measurements and as complementary search
of the new physics effects beyond the SM. Particle factories at the
energy range (3-11 GeV) with a luminosity range $L=10^{33}-10^{36}$
cm$^{-2}$ s$^{-1}$ are devoted to detailed studies of heavy-flavored
states like charm hadrons, tau lepton and bottom hadrons, \cite{Bona:2007qt,Kurokawa:2001nw,Aubert:2001tu}.
These particles are of special interest for several reasons: i.e.
quantitative tests of QCD at the interface between the perturbative
and non-perturbative regimes, matter-antimatter asymmetry, origin
of the CP violation, precision studies of tau and its neutrino and
new physics effects.

The tau-charm factory has been devised to investigate the properties
of the charm physics and tau leptons (which are the heaviest and last
discovered family of the lepton sector of the SM). The tau lepton
offers some unique properties which make it an excellent tool for
challenging our current understanding of particle physics. The charm-tau
factory operates at 3-6 GeV center of mass energy. The tau physics
have many significant properties which give motivations to analyze
the interactions of the tau leptons (for recent reviews on tau physics,
see for example \cite{Lafferty:2009zz,Pich:2007cu,Banerjee:2007is,Marciano:2008zz}).
A high luminosity of $L=10^{35}$ cm$^{-2}$ s$^{-1}$ charm-tau factory
of the Turkish Accelerator Center (TAC) has been proposed \cite{tac}.
With this high luminosity of the factory it is promising to search
for the new physics effects in flavour physics.

Amidst the other beyond the SM scenarios, the unparticle physics recently
proposed by Georgi have very interesting pecularities, \cite{Georgi:2007ek},
and can be investigated through $e^{+}e^{-}\to\tau^{+}\tau^{-}$ process
at a tau factory. The unparticle physics has been proposed with the
possibility of new physics effects of a hidden scale invariant sector
living at a very high energy scale on our low energy physics observables.
According to this argument, if the scale invariance occurs in the
nature, there may be a scale invariant sector at a very high energy
scale, and at those energy scales both a scale invariant sector presented
by a set of the Banks-Zaks oparators ${\cal O}_{BZ}$ and the Standard
Model(SM) operators ${\cal O}_{SM}$ can be coexisted \cite{Banks:1981nn}.
Based on the notation of \cite{Georgi:2007ek}, these two sets of
operators can interact through the exchange of particles with a mass
scale ${\cal M}_{{\cal U}}^{k}$ in the following form

\begin{eqnarray}
\frac{1}{{\cal M}_{{\cal U}}^{k}}{O}_{BZ}{O}_{SM}\label{eq:1}\end{eqnarray}
where $k>0$, the SM and BZ operators are defined as ${O}_{SM}\in{\cal O}_{SM}$
with mass dimension $d_{SM}$ and ${O}_{BZ}\in{\cal O}_{BZ}$ with
mass dimension $d_{BZ}$. At an energy scale $\Lambda_{{\cal U}}$,
the renormalizable couplings of ${\cal O}_{BZ}$ imply a dimensional
transmutation in the BZ-sector such that below $\Lambda_{{\cal U}}$
the BZ-operators correspond to the so called unparticle operators.
Thus, after the dimensional transmutation (\ref{eq:3-1}) can be written
as

\begin{eqnarray}
\frac{C_{{\cal U}}\Lambda_{{\cal U}}^{d_{BZ}-d}}{{\cal M}_{{\cal U}}^{k}}{O}_{{\cal U}}{O}_{SM}\label{eq:2}\end{eqnarray}
where $d$ is the scaling mass dimension of the unparticle operator
$O_{{\cal U}}$, and the constant $C_{{\cal U}}$ is a coeficient
function.

There have been numerious works on the implications of the unparticle
physics, for example one can consult \cite{Georgi:2007ek,Georgi:2007si,Cheung:2007ue,Bander:2007nd}
and references therein. In the scope of the present work, i.e. unparticle
effects on tau/charm physics, the unparticle effects on particle and
antiparticle oscillations in meson-antimeson has been investigated
by \cite{Chen:2007cz}, \cite{Parry:2008sr}, and \cite{Wei:2008zzc},
and tau decays by \cite{Hektor:2008xu}. In their analysis they study
both charm and B-physics. In this work, we take into account the tau
physics option of the tau/charm factories and we study on the possible
unparticle signatures through $e^{+}e^{-}\to\tau^{+}\tau^{-}$ process.

\section{Tau Pair Production}

\begin{figure}
\includegraphics[scale=0.6]{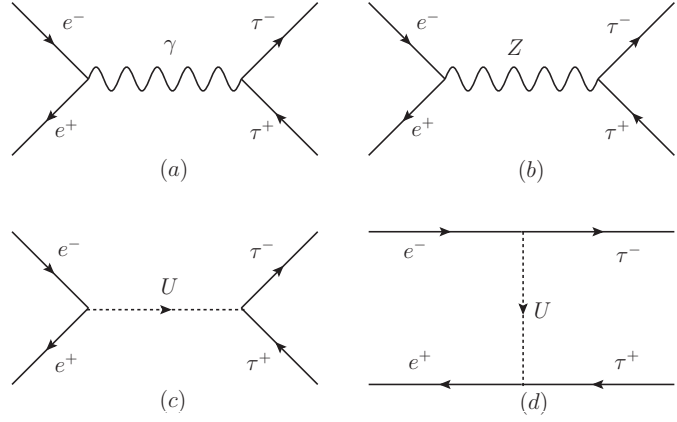} 

\caption{Feynman diagrams for the tree level contributions of SM and unparticles
to $e^{+}e^{-}\to\tau^{+}\tau^{-}$ process. \label{fig:fig1}}

\end{figure}

\begin{figure}
\includegraphics{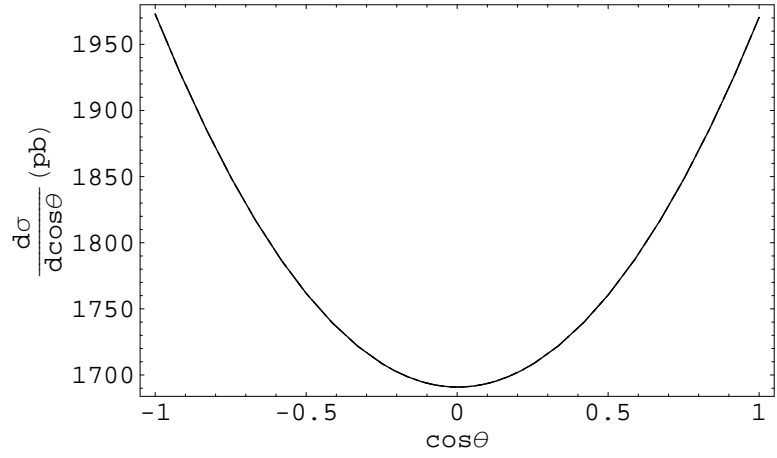} 

\caption{Differential cross section for the angular distribution of the tau
lepton due to the SM contribution at $\sqrt{s}=4.2$ GeV.\label{fig:fig2}}

\end{figure}

\begin{figure}
\includegraphics{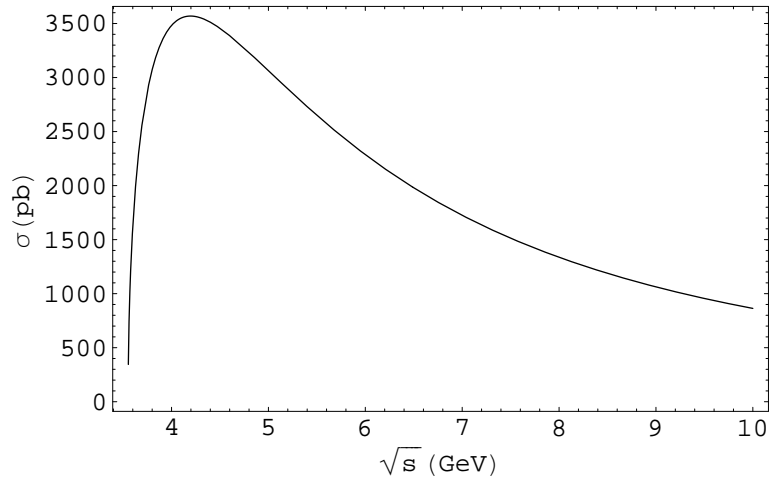} 

\caption{Total cross section due to the SM contribution depending on the center
of mass energy.\label{fig:fig3}}

\end{figure}

The optimum energy for studying a particular particle at an $e^{+}e^{-}$
collider correspond to the region near its pair production threshold,
which generally has the highest cross sections, lowest background
and other favorable experimental conditions. In the framework of the
SM, tau-lepton pair production occurs through the exchange of photon
and $Z$-boson in the $s$-channel as shown in Fig. \ref{fig:fig1}(a-b).
The differential cross section for the scattering process $e^{+}e^{-}\to\tau^{+}\tau^{-}$
keeping the terms including the mass of the tau lepton is given by

\begin{eqnarray}
\frac{d\sigma}{dt}= &  & \frac{g_{e}^{4}}{16\pi s^{2}}\Big\{\frac{2(2m_{\tau}^{4}+s^{2}-4m_{\tau}^{2}t+2st+2t^{2})}{s^{2}}+\frac{c_{a}^{2}c_{v}^{2}(2m_{\tau}^{4}+3s^{2}+6st+2t^{2}-2m_{\tau}^{2}(3s+2t))}{4c_{w}^{4}s_{w}^{4}[(s-m_{z}^{2})^{2}+m_{z}^{2}\Gamma_{z}^{2}]}\nonumber \\
 &  & +\frac{c_{v}^{4}(2m_{\tau}^{4}+s^{2}-4m_{\tau}^{2}t+2st+2t^{2})+c_{a}^{4}(2m_{\tau}^{4}+s^{2}+2st+2t^{2}-4m_{\tau}^{2}(s+t))}{8c_{w}^{4}s_{w}^{4}[(s-m_{z}^{2})^{2}+m_{z}^{2}\Gamma_{z}^{2}]}\nonumber \\
 &  & +\frac{(s-m_{z}^{2})[c_{v}^{2}(2m_{\tau}^{4}+s^{2}-4m_{\tau}^{2}t+2st+2t^{2})+c_{a}^{2}s(s-2m_{\tau}^{2}+2t)]}{c_{w}^{2}s_{w}^{2}s[(s-m_{z}^{2})^{2}+m_{z}^{2}\Gamma_{z}^{2}]}\Big\}\label{eq:3-1}\end{eqnarray}
where $g_{e}=\sqrt{4\pi\alpha}$ is the electromagnetic coupling constant;
$c_{w}$ and $s_{w}$ are the cosine and sine of the weak mixing angle
$\theta_{w}$. Here $c_{v}$ and $c_{a}$ are the vector and axial-vector
couplings of the $Z$-boson; $m_{\tau}$ is the tau lepton mass; $m_{z}$
and $\Gamma_{z}$ denote the $Z$-boson mass and decay width, respectively
(for numerical values see \cite{Yao:2006px}). The Mandelstam variables
$s=(p_{e^{-}}+p_{e^{+}})^{2}=(p_{\tau^{+}}+p_{\tau^{-}})^{2}=4E_{e+}E_{e-}$,
and $t=m_{\tau}^{2}-s(1-\beta\cos\theta)/2$ with $\beta=\sqrt{1-4m_{\tau}^{2}/s}$
and $\cos\theta=p_{z}/|\vec{p}|$. The differential cross section
and total cross section are presented in Fig. \ref{fig:fig2} and
\ref{fig:fig3}. For $s\ll m_{z}^{2}$ the total cross section is
found to be $\sigma\approx2\pi\alpha^{2}\beta(3-\beta^{2})/3s$. The
pair production threshold is around 3.5 GeV while the total cross
section shows a peak around 4.2 GeV with $\sim3.6$ nb. From Figure
\ref{fig:fig3} we find the cross section $\sim3.1$ nb at 3.8 GeV,
which is about 3.5 times larger than that at 10 GeV.

The effective interaction among the SM fermions and the scalar unparticles
is given by \begin{eqnarray}
\frac{\lambda_{\alpha\beta}}{\Lambda_{U}^{d-1}}\bar{f_{\alpha}}f_{\beta}O_{{\cal U}}+h.c.\label{eq:3}\end{eqnarray}
where $\alpha,\beta$ stand for the fermion flavors. Therefore, the
contributions from the exchange of the scalar unparticles to the scattering
amplitude for the process $e^{+}e^{-}\to\tau^{+}\tau^{-}$, Fig. \ref{fig:fig1}(c-d),
are given

\begin{eqnarray}
M_{{\cal U}}^{s}= &  & [f_{d}(\lambda_{0})][\bar{v}_{e^{+}}(p_{2})u_{e^{-}}(p_{1})][\bar{u}_{\tau^{-}}(p_{3})v_{\tau^{+}}(p_{4})][s]^{d-2}e^{id\pi}\label{4}\\
M_{{\cal U}}^{t}= &  & [f_{d}(\lambda_{0}')][\bar{u}_{\tau^{-}}(p_{3})u_{e^{-}}(p_{1})][\bar{v}_{e^{+}}(p_{2})v_{\tau^{+}}(p_{4})][-t]^{d-2}\end{eqnarray}
where

\begin{eqnarray}
f_{d}(\lambda_{0})=\Big[\frac{\lambda_{0}^{2}}{\Lambda_{{\cal U}}^{2d-2}}\Big]\Big[\frac{A_{d}}{2\sin d\pi}\Big]\label{5}\end{eqnarray}
with

\begin{eqnarray}
A_{d}=\frac{16\pi^{5/2}}{(2\pi)^{2d}}\frac{\Gamma(d+1/2)}{\Gamma(d-1)\Gamma(2d)}\label{6}\end{eqnarray}

In Eqs. (\ref{4} - \ref{5}) we take into account the possible lepton
flavour conserving ($\lambda_{ee}\equiv\lambda_{\tau\tau}\equiv\lambda_{0}$),
and flavour changing ($\lambda_{e\tau}\equiv\lambda_{0}'$) couplings
for the unparticle interactions.

The differential cross section for both SM and unparticle contributions
can be found as follows

\begin{eqnarray*}
\frac{d\sigma}{dt}=\frac{[|M_{SM}|^{2}+|M_{U}^{s}|^{2}+|M_{U}^{t}|^{2}+2Re[M_{U}^{t*}M_{U}^{s}]]}{16\pi s^{2}}\end{eqnarray*}
where there is no interference between the SM and the scalar unparticle
contributions(here we neglect the electron mass), and one can easily
find the amplitudes due to the unparticle exchanges

\begin{eqnarray}
|M_{U}^{s}|^{2} & = & [f_{d}(\lambda_{0})]^{2}[s]^{2d-2}[1-4m_{\tau}^{2}/s]\\
|M_{U}^{t}|^{2} & = & [f_{d}(\lambda_{0}')]^{2}[-t]^{2d-2}[1-m_{\tau}^{2}/t]^{2}\\
2Re[M^{t}{}_{U}^{*}M_{U}^{s}] & =- & [f_{d}(\lambda_{0})f_{d}(\lambda_{0}')][s]^{d-1}[-t]^{d-2}[t+m_{\tau}^{2}]\cos d\pi\end{eqnarray}

\begin{figure}
\includegraphics{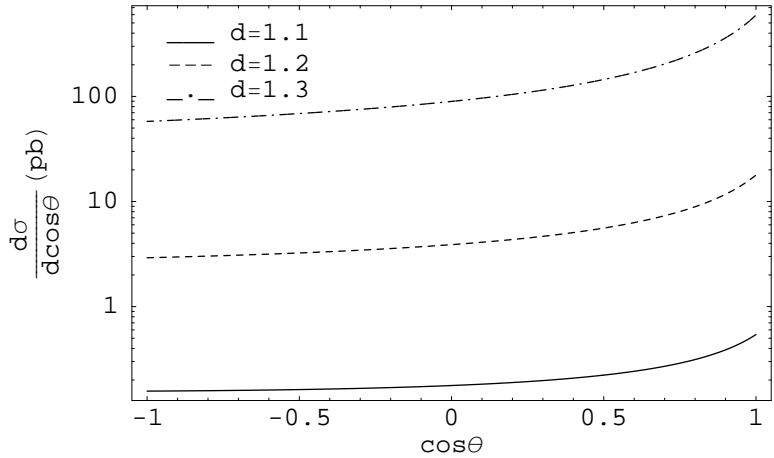} 

\caption{The angular distributions for unparticle contribution at $\sqrt{s}$=4.2
GeV. Here, we assume $\lambda_{0}=\lambda_{0}'=0.3$ and $\Lambda_{{\cal U}}=1000$
GeV.\label{fig:fig4}}

\end{figure}

\begin{figure}
\includegraphics{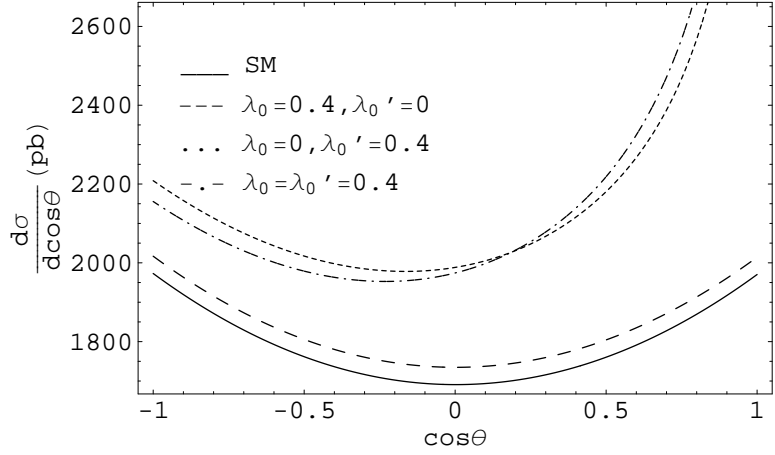} 

\caption{The SM and unparticle contributed differential cross sections at $\sqrt{s}$=4.2
GeV. For the unparticle contribution, we assume $d=1.1$ and $\Lambda_{{\cal U}}=1000$
GeV.\label{fig:fig5}}

\end{figure}

\begin{figure}
\includegraphics{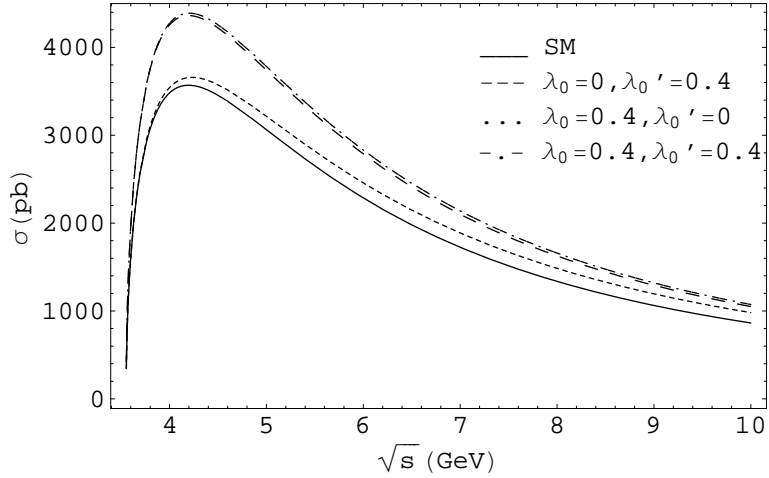} 

\caption{The SM cross section and the unparticle contributed cross sections
with respect to the center of mass energy of the collider. For the
unparticle contribution, we assume $d=1.1$, and $\Lambda_{{\cal U}}=1000$
GeV.\label{fig:fig6} }

\end{figure}

\begin{figure}
\includegraphics{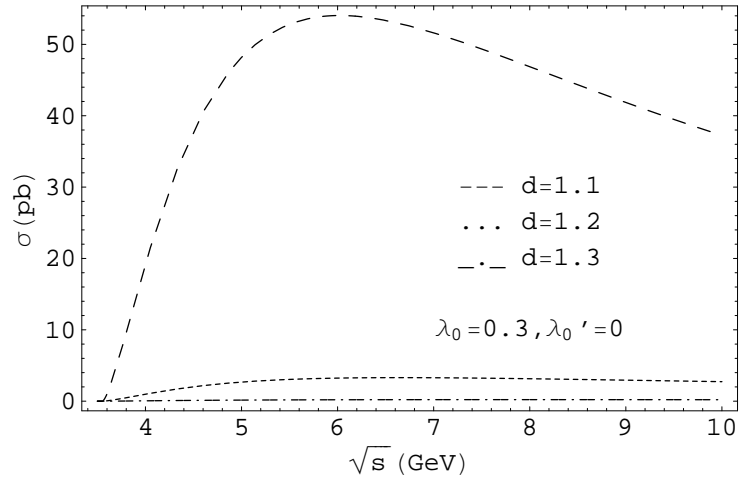} 

\caption{The cross section for pure unparticle contribution with flavor conserving
couplings depending on the center of mass energies for different scaling
mass dimension $d=1.1$, 1.2 and 1.3. We assume couplings $\lambda_{0}=0.3$,$\lambda_{0}'=0$
and $\Lambda_{{\cal U}}=1000$ GeV.\label{fig:fig7}}

\end{figure}

\begin{figure}
\includegraphics{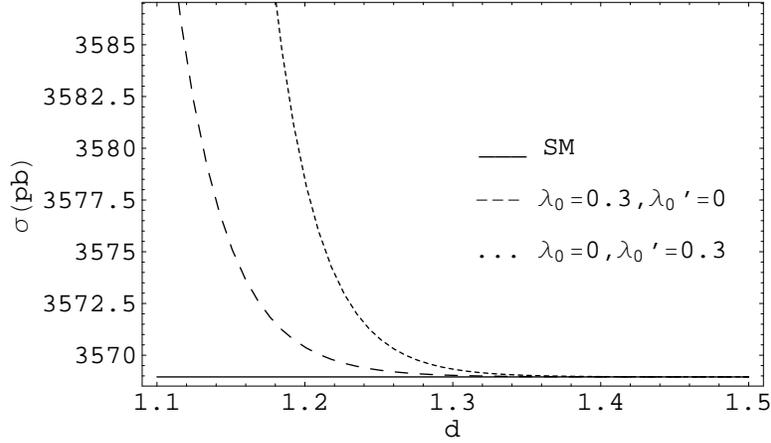} 

\caption{Dependence of the unparticle contributed total cross section on d
for different coupling configurations at $4.2$ GeV. We assume $\Lambda_{{\cal U}}=1000$
GeV.\label{fig:fig8}}

\end{figure}

In the Figure \ref{fig:fig4}, we plot the pure unparticle contributed
differential cross sections with respect to $\cos\theta$ for different
values of the scaling parameter $d$, where $\theta$ is the scattering
angle of the tau lepton in the center of mass frame. In Fig. \ref{fig:fig5},
for an illustrution we show the effects from the flavor preserving
and the flavor non-preserving unparticle couplings, here we assume
three possible configuration $(\lambda=0.4,\lambda^{\prime}=0.),(\lambda=0.,\lambda^{\prime}=0.4),(\lambda=\lambda^{\prime}=0.4)$
for $d=1.1$, where we take $\Lambda_{{\cal U}}=1000$ GeV. In Fig.
\ref{fig:fig6}, we plot total cross section depending on the center
of mass energies in the range 3.5-10 GeV. From this figure one can
see the effects of the flavor preserving and the flavor non-preserving
unparticle couplings, which could be measured at future factories.
Fig. \ref{fig:fig8} shows the cross section depending on the scaling
dimension $d$ for different coupling configurations. From those figures,
one should notice that the effects of the unparticle, for the given
parameter configurations, can be discriminated from the SM background
for the values $d<1.4$. To compare the contributions we give the
numerical values of the total cross sections with and without unparticle
effects in Table~\ref{tab:tab1}. The contributions of the scalar
unparticle with flavour conserving couplings are shown in Fig. \ref{fig:fig7},
for different scaling dimension $d$, depending on the center of mass
energies between $3.5-10$ GeV. 

\begin{table}
\caption{The cross sections for the SM and unparticle contributions for $d=1.1$
and $\Lambda=1000$ GeV at different energies. \label{tab:tab1}}

\begin{tabular}{|c|c|c|c|c|}
\hline 
\multicolumn{1}{|c|}{} & \multicolumn{4}{c|}{$\sigma$(nb)}\tabularnewline
\hline 
$\sqrt{s}$(GeV)  & SM+U($\lambda_{0}=\lambda_{0}'=0.3$)  & SM+U($\lambda_{0}=0.3$)  & SM+U($\lambda_{0}'=0.3$)  & SM\tabularnewline
\hline 
3.8  & 3.291  & 3.082  & 3.288  & 3.072\tabularnewline
\hline 
4.2  & 3.828  & 3.596  & 3.819  & 3.568\tabularnewline
\hline 
10.0  & 0.929  & 0.900  & 0.922  & 0.863\tabularnewline
\hline
\end{tabular}
\end{table}

\section{Limits and Discussion}

\begin{figure}
\includegraphics{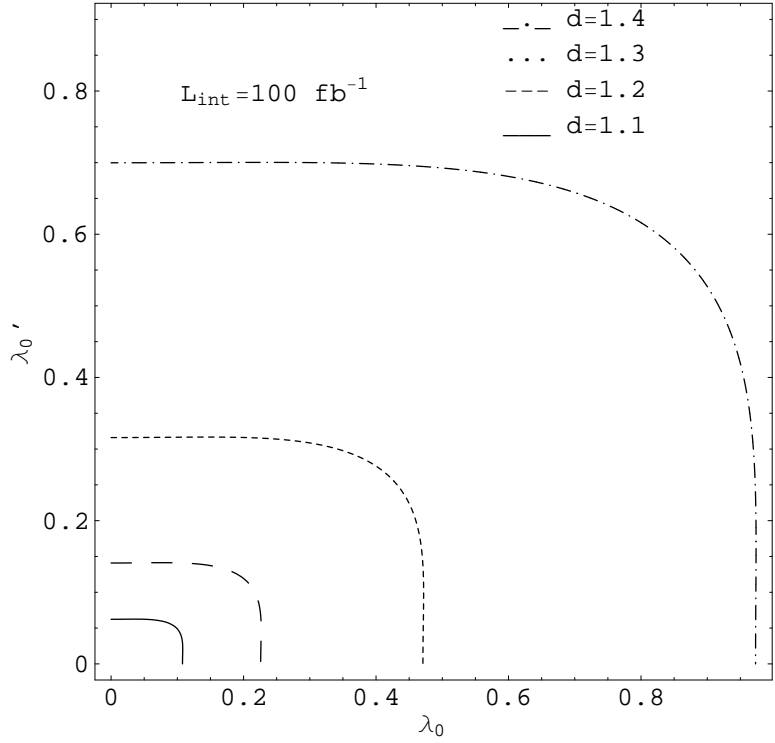} 

\caption{Contour plot to represent the $95\%$C.L. limits on the unparticle
couplings for ${\cal L}_{int}=100$ fb$^{-1}$/yr at 4.2 GeV. We take
$\Lambda_{{\cal U}}=1000$ GeV.\label{fig:fig9}}

\end{figure}

\begin{figure}
\includegraphics{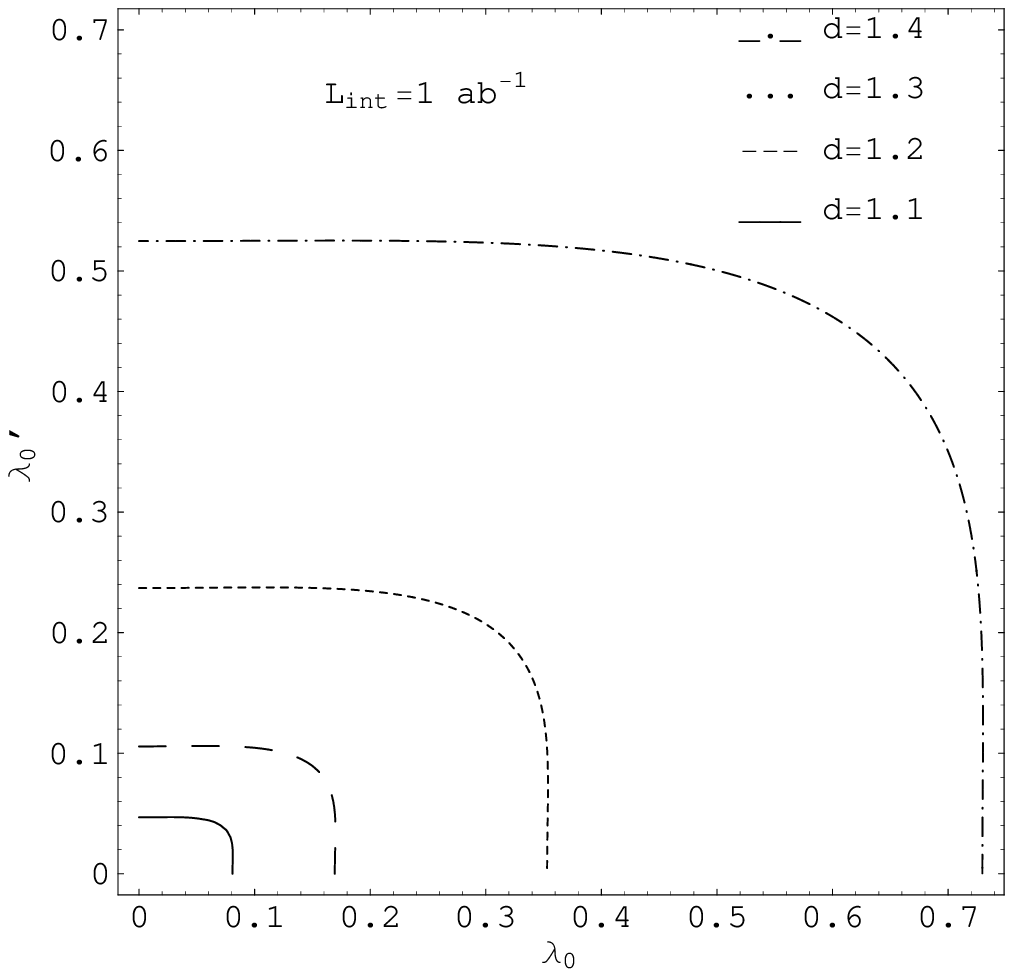} 

\caption{Contour plot to represent the $95\%$C.L. limits on the unparticle
couplings for ${\cal L}_{int}=1000$ fb$^{-1}$/yr at 4.2 GeV. We
take $\Lambda_{{\cal U}}=1000$ GeV.\label{fig:fig10}}

\end{figure}

To calculate the limits on the unparticle couplings we use the standard
two-parameter chi-square analysis for the chi-square function

\begin{eqnarray}
\chi^{2}=\frac{[\sigma_{SM+U}(\lambda_{0},\lambda_{1})-\sigma_{SM}]^{2}}{(\Delta\sigma_{SM})^{2}}\end{eqnarray}
with

\begin{eqnarray}
{\Delta\sigma_{SM}=\sigma_{SM}\sqrt{[(\frac{1}{\sqrt{N}})^{2}+\delta_{syst}^{2}]}}\\
N=\sigma_{SM}.{\cal L}_{int}.\epsilon\end{eqnarray}
where $N$ is the number of events, ${\cal L}_{int}$ is the collider
integrated luminosity, and $\epsilon$ is the efficiency for the interested
channel. For the {95\%}C.L. we extract upper limits on the couplings
assuming $\chi^{2}=5.99$ for two parameters. Assuming the optimal
integrated luminosity ${\cal L}_{int}=100$ fb$^{-1}$/yr and ${\cal L}_{int}=1$
ab$^{-1}$/yr at $4.2$ GeV, we present the limits on the flavor preserving
and the flavor violating couplings in the contour plot, Figures \ref{fig:fig9}
and \ref{fig:fig10}. The limits are listed in the Table~\ref{tab:tab2}
for different center of mass energies available at the flavor factories. 

\begin{table}[!htp]
\caption{Upper limits from the $\chi^{2}$ analysis on the couplings $(\lambda_{0},\lambda_{0}^{\prime})$
for $\Lambda_{{\cal U}}=1000$ GeV, with the ${\cal L}_{int}=100$
fb$^{-1}$/yr and 1 ab$^{-1}$/yr according to the scaling mass dimension
$d$ at the center of mass energies $3.8$, $4.2$ and 10 GeV. }

\label{tab:tab2} \centering{}\begin{tabular}{|c|c|c|c|c|c|}
\hline 
$\sqrt{s}$(GeV) & Luminosity & $d=1.1$  & $d=1.2$  & $d=1.3$  & $d=1.4$ \tabularnewline
\hline
$3.8$ & 100 fb$^{-1}$ & (0.131,0.059) & (0.276,0.136) & (0.581,0.308) & (1.210,0.690)\tabularnewline
\hline 
 & 1 ab$^{-1}$  & (0.098,0.045) & (0.207,0.102) & (0.436,0.231) & (0.910,0.518)\tabularnewline
\hline
\hline 
$4.2$ & 100 fb$^{-1}$ & (0.102,0.059) & (0.214,0.133) & (0.445,0.299) & (0.920,0.662)\tabularnewline
\hline 
 & 1 ab$^{-1}$  & (0.076,0.044) & (0.160,0.099) & (0.334,0.224) & (0.690,0.496)\tabularnewline
\hline
\hline 
$10$ & 100 fb$^{-1}$ & (0.079,0.071) & (0.152,0.149) & (0.291,0.303) & (0.552,0.601)\tabularnewline
\hline 
 & 1 ab$^{-1}$  & (0.059,0.053) & (0.114,0.112) & (0.219,0.227) & (0.414,0.451)\tabularnewline
\hline
\end{tabular}
\end{table}

To conclude, in this work we have studied the unparticle effects on
tau pair production at low energy particle factories. We analyze in
detail the limits on the flavor conserving/changing scalar unparticle
couplings at the collider energies $\sqrt{s}$=3.8, 4.2 and 10 GeV
with the luminosities of 100 fb$^{-1}$/yr and $1$ ab$^{-1}$/yr.
We find the limits for $d<1.4$ assuming the unparticle energy scale
$\Lambda_{{\cal U}}=1000$ GeV are consistent with the limits found
in the literature. We would like to remark that higher center of mass
energy has an advantage for the measurements of flavor conserving
unparticle couplings. However, the flavor changing unparticle couplings
for $d<1.3$ can be measured more accurately at a charm/tau factory.
\begin{acknowledgments}
This work is supported by the State Planning Organization (DPT) with
grant number DPT2006K-120470. \end{acknowledgments}

\end{document}